\documentclass[prl,nofootinbib,amsmath,twocolumn,notitlepage,preprintnumbers,floatfix]{revtex4-1}
\usepackage{amssymb,esvect,amsmath,graphicx,latexsym,amsthm,slashed,eso-pic, tabularx, bbold}
\usepackage{xfrac}
\usepackage{enumitem}
\usepackage{amssymb, amsmath}
\usepackage[colorlinks=true,urlcolor=blue
,anchorcolor=blue,citecolor=blue,filecolor=blue,linkcolor=blue,menucolor=blue
]{hyperref}

\newcommand{\beq}{\begin{equation}} \newcommand{\eeq}{\end{equation}}
\newcommand{\bea}{\begin{eqnarray}} \newcommand{\eea}{\end{eqnarray}}

\def\lsim{\mathrel{\raise.3ex\hbox{$<$\kern-.75em\lower1ex\hbox{$\sim$}}}}
\def\gsim{\mathrel{\raise.3ex\hbox{$>$\kern-.75em\lower1ex\hbox{$\sim$}}}}

\newcommand{\Eq}[1]{Eq.~(\ref{#1})} \newcommand{\Eqs}[2]{Eqs.~(\ref{#1}) and (\ref{#2})}

\newcommand{\be}{\begin{eqnarray}}
\newcommand{\ee}{\end{eqnarray}}

\newcommand{\benum}{\begin{enumerate}}
\newcommand{\eenum}{\end{enumerate}}
\newcommand{\bi}{\begin{itemize}}
\newcommand{\ei}{\end{itemize}}

 \newcommand{\brac}[2]{ \left( \frac{#1}{#2} \right) }

\usepackage{hyperref}
\usepackage[mathlines]{lineno}

\begin{document}

\preprint{FERMILAB-PUB-21-344-T}
\title{Towards a Realistic Model of Dark Atoms to Resolve the Hubble Tension}

\author{Nikita Blinov$^{a,b}$}
\thanks{nblinov@uvic.ca, ORCID: \href{http://orcid.org/0000-0002-2845-961X}{http://orcid.org/0000-0002-
2845-961X}; now at Department of Physics and Astronomy, University of Victoria, Victoria, BC V8P 5C2, Canada}

\author{Gordan Krnjaic$^{a,b,c}$}
\thanks{krnjaicg@fnal.gov, ORCID: \href{http://orcid.org/0000-0001-7420-9577}{http://orcid.org/0000-0001-7420-9577}}

\author{Shirley Weishi Li $^{a}$}
\thanks{shirleyl@fnal.gov, ORCID: \href{http://orcid.org/0000-0002-2157-8982}{http://orcid.org/0000-0002-2157-8982}}

\affiliation{$^a$Fermi National Accelerator Laboratory, Theoretical Physics Department}
\affiliation{$^b$University of Chicago, Kavli Institute for Cosmological Physics}
\affiliation{$^c$University of Chicago, Department of Astronomy and Astrophysics}

\date{\today}

\begin{abstract}
It has recently been shown that a subdominant hidden sector of 
atomic dark matter in the early universe provides a novel avenue towards resolving the Hubble ($H_0$) tension while maintaining good agreement
with Cosmic Microwave Background era observables. 
However, such a mechanism 
 requires a hidden sector whose energy density ratios are the same as in our sector and whose 
 recombination also takes place at redshift $z \approx 1100$, 
 which presents an apparent
 fine tuning. We introduce a realistic model of this scenario that dynamically enforces
 these coincidences without fine tuning. In our setup, the hidden sector
 contains an identical copy of Standard Model (SM) fields, but has a
 smaller Higgs vacuum expectation value (VEV) and a lower temperature. 
The baryon asymmetries and reheating temperatures
in both sectors arise from the decays of an Affleck-Dine scalar field, whose branching ratios automatically ensure that the reheating temperature in each sector is proportional to the corresponding Higgs VEV. 
The same setup also naturally ensures that the Hydrogen binding energy in each sector is proportional to the corresponding VEV, so the ratios of binding energy to temperature are approximately equal in the two sectors. Furthermore, our scenario predicts a correlation between 
the SM/hidden temperature ratio and the atomic dark matter abundance and automatically yields values for these quantities 
favored by concordant early- and late-universe measurements of $H_0$.
\end{abstract}

\maketitle

\section{Introduction}

The longstanding tension between the early universe \cite{Planck2020} and local \cite{Riess_2021} extractions of the
Hubble constant $H_0$ may signal the breakdown of the standard $\Lambda$ Cold Dark Matter ($\Lambda$CDM)
 paradigm (see \cite{Verde:2019ivm} for reviews). Although
recent local measurements using the tip of the red giant branch method suggest that 
late time measurements might be more compatible with early universe extractions \cite{Freedman:2021ahq}, 
it remains to be seen whether these results will ultimately converge without 
the need for new physics \cite{Anand:2021sum}. 
While the tension may still be an artifact of systematic error in at least one measurement technique,
many models of new physics have been proposed to resolve the discrepancy (see \cite{Knox:2019rjx,DiValentino:2021izs,Schoneberg:2021qvd}
for recent reviews).

It has recently been shown that
a subdominant component of atomic dark matter (ADM) \cite{Foot:2002iy,Foot:2003jt,Kaplan:2009de,Kaplan:2011yj,Cline_2012,Cyr_Racine_2013,Cline:2013pca,Cline:2021itd} 
can viably increase the early universe value of $H_0$ while
preserving a good fit to other cosmic microwave background (CMB) and
baryon acoustic oscillation (BAO) observables \cite{Cyr-Racine:2021alc}. In
this scenario, the usual $\Lambda$CDM model is supplemented with a hidden sector of 
dark atoms, photons and neutrinos; the former accounts for a few percent of 
the dark matter and the latter contribute to the radiation density during the CMB era. Unlike
other models that propose modifications to the early universe, this scenario mimics 
the behavior of visible matter by maintaining the same matter/radiation ratio and 
undergoing recombination at $z \approx 1100$. 
The addition of this sector approximately mimics the
scaling symmetry 
\be
\label{sym}
\rho_i \to f^2 \!  \rho_i~~,~~ \sigma_T n_e  \to f  \sigma_T n_e ~~,~~
A_s \to A_s f^{1-n_s },~~~
\ee
where $\rho_i$ is the $i^\text{th}$ energy density component, $n_e$ is the 
electron density, $\sigma_T = 8\pi \alpha^2/(3m_e^2)$ is the Thomson cross section, $A_s$ is the amplitude of 
scalar fluctuations, $n_s$ is the spectral tilt, and $f$ is an 
arbitrary constant. The transformation in \Eq{sym} preserves
the form of all cosmological perturbation equations in linear theory, thereby retaining the good agreement of CMB/BAO 
predictions with data \cite{Cyr-Racine:2021alc}. 

As noted in Ref. \cite{Cyr-Racine:2021alc}, this framework presents
two main observational challenges: 1) the model favors a small 
CMB value of the cosmological helium fraction, $Y_p \approx 0.17$ to within a few percent, which is in clear tension
with the consensus value from Big Bang Nucleosynthesis (BBN) $Y_p = 0.245 \pm 0.003$ \cite{ParticleDataGroup:2020ssz};
and 2) the best fit hidden/visible temperature ratio satisfies $T^\prime/T \approx 0.7$,
corresponding to a large value of $\Delta N_{\rm eff} \approx 1.6$ during BBN, assuming identical
SM field content in the hidden sector. 
Thus the ADM mechanism shifts the tension into parameters that 
affect a different era of cosmological history. 

Furthermore, at face value, mimicking this approximate symmetry in \Eq{sym}
 requires an ad-hoc coincidence to ensure that dark and visible 
 recombination {\it both} occur at $z \approx 1100$. Since the reheating temperature in each sector is an initial condition and 
 the Hydrogen binding energy depends on strong, electromagnetic, and Higgs couplings, 
 such a coincidence across sectors with different masses and thermal histories seems extremely 
 unlikely at first glance; this situation calls for a dynamical explanation.

 In this paper we show how such a coincidence can arise
in a realistic model that accounts for the full
cosmological history of the atomic hidden sector. We model the hidden sector
as an identical copy of the Standard Model with lighter elementary particles and derive the baryon asymmetry
and initial temperature in each sector through its coupling to an Affleck-Dine 
scalar field that dominates the early, post-inflationary universe. 

Our model preserves all of the beneficial features identified in Ref. \cite{Cyr-Racine:2021alc} while eliminating
fine tuning needed to time hidden sector recombination. Furthermore, this scenario correlates 
the interacting DM fraction $f_{\rm adm} \equiv \Omega_{\rm adm}/\Omega_{\rm cdm}$ and the $T^\prime/T$ ratio,
thereby removing one free parameter to make our scenario more predictive. 
However, since we recover the same hidden sector studied in
 Ref. \cite{Cyr-Racine:2021alc}, it inherits the 
 tension in $\Delta N_{\rm eff}$ and $Y_p$.

\section{Model Overview}
Inspired by Twin Higgs models \cite{Chacko:2005pe}, we postulate a mirror 
hidden sector which contains an identical copy of all Standard Model 
(SM) fields, coupling constants, and gauge interactions. As in Twin
Higgs models, the two sectors here have different values for the 
Higgs vacuum expectation values (VEVs), but in our scenario 
we demand that $v^\prime/v < 1$ where
$v^{(\prime)}$ is the SM (hidden) Higgs VEV and we use primed symbols to refer to 
hidden sector quantities throughout this work. Since all other
couplings are identical, the QCD confinement scales in
both sectors satisfy $\Lambda_{\rm qcd} \approx \Lambda_{\rm qcd}^\prime \approx 200$ MeV
\cite{Chacko:2018vss}, which also yields a similar proton mass for both sectors; all other
elementary particle masses in the hidden sector are scaled down by an overall factor of $v^\prime/v$
relative to the SM. 
Unlike in Twin Higgs models, our setup does not invoke any direct couplings between the two
sectors and does not address the electroweak hierarchy problem, so there are no collider constraints
on the VEV ratio.

To ensure viable bound state formation to explain the Hubble tension in this framework, we must
satisfy the following conditions: 
\begin{enumerate}
\item  There must be a hidden baryon asymmetry to prevent hidden sector particles from completely annihilating into radiation.
 \item Any interaction between sectors must be sufficiently 
 feeble to prevent them from reaching thermal equilibrium.
 
  \item The reheating temperature in each sector must be directly proportional to the corresponding Higgs VEV. Since we have exhausted all the freedom in choosing field content and coupling constants, the VEV dependence in the
Hydrogen binding energy $B \propto v$ is compensated by $T\propto v$, so $B/T \approx B^\prime/T^\prime$ 
and both sectors undergo recombination at the same time.
 \end{enumerate}
 
 In what follows, we realize all of these requirements by coupling both sectors
 to an Affleck-Dine field whose decays simultaneously yield the requisite particle asymmetries and 
 temperature relations to reconcile local and CMB measurements of $H_0$ with a subdominant atomic dark sector;
we assume the remaining CDM in this scenario arises from a different source.

While our model and the phenomenological scenario in Ref. \cite{Cyr-Racine:2021alc} both
approximately realize the symmetry in \Eq{sym}, we add one additional source of 
scaling violation in the hidden sector Thomson cross section $\sigma^\prime_T/\sigma_T = (v/v^\prime)^2 > 1$.
The hidden radiation is more tightly coupled to its matter content as a result. Nonetheless, we find that this additional 
symmetry violating detail does not spoil the good CMB/BAO fit from Ref. \cite{Cyr-Racine:2021alc}.

\section{Cosmological Evolution}

 \begin{figure}
 \includegraphics[width=3in,angle=0]{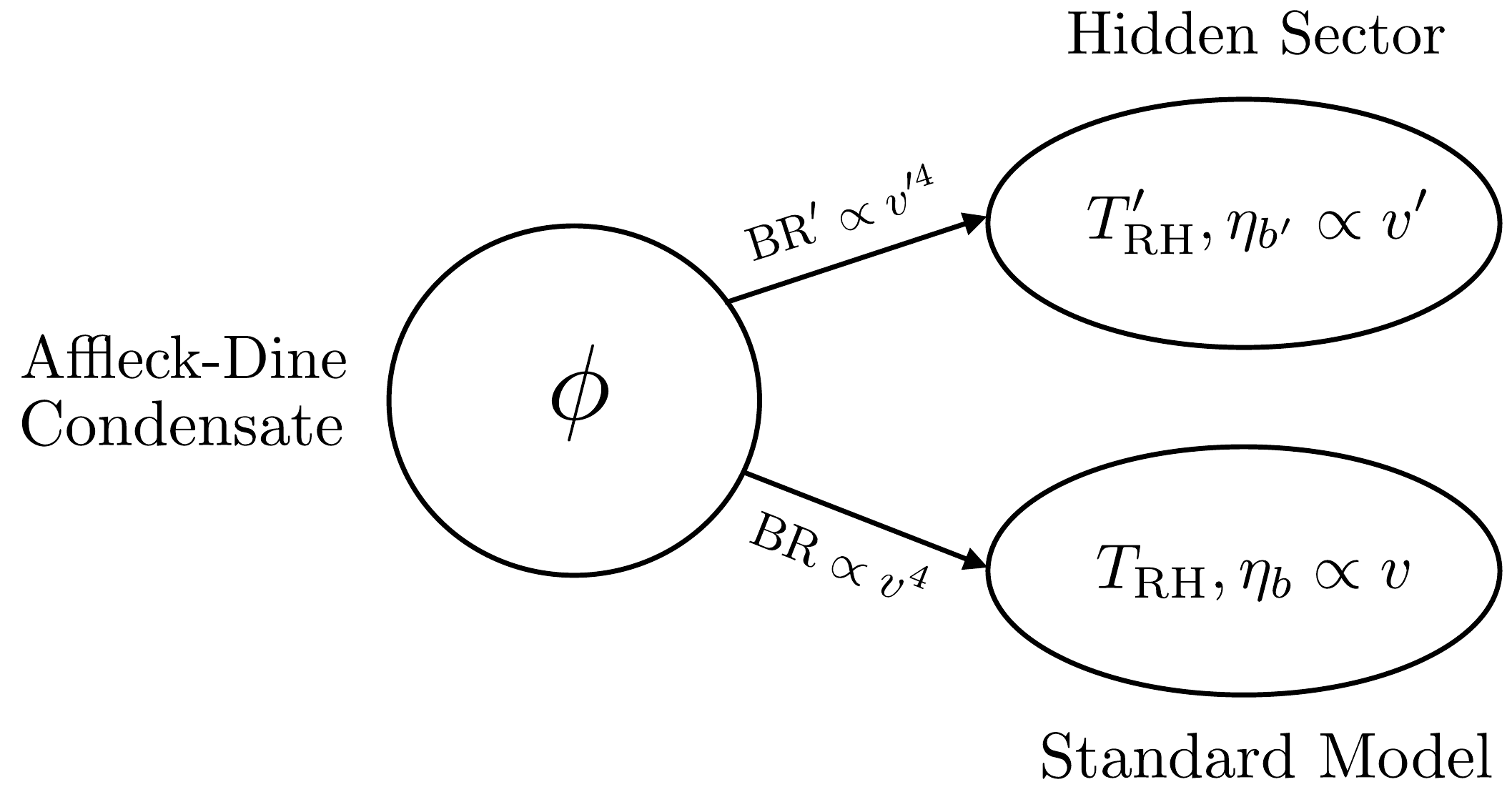}
\caption{ 
Schematic illustration of our setup. A non-relativistic condensate of 
Affleck-Dine scalar field $\phi$ dominates the energy
density of the post-inflationary early universe and carries net baryon number.
 Upon decay, $\phi$ transfers its asymmetry to SM and hidden sector fields with
 branching ratios proportional to the Higgs VEV in each sector. 
}
\label{cartoon}
\end{figure}

We assume the post-inflationary early universe is dominated by a complex 
scalar field $\phi = r e^{i\theta}/\sqrt{2}$ that carries baryon number $B_\phi$ and 
realizes Affleck-Dine baryogenesis \cite{Affleck:1984fy} as depicted schematically 
in Fig. \ref{cartoon}. In polar coordinates, the
scalar potential is 
\be
\label{pot}
V(r,\theta) = \frac{m_\phi^2}{2} r^2 + \frac{\lambda}{8} r^4
- \frac{\kappa}{8} r^4 \cos4 \theta ~~,
\ee
where $m$ is the $\phi$ mass, $|\kappa| \ll  |\lambda|$ are dimensionless couplings,
 and the explicit $\theta$ dependence in the last term provides a source of baryon number violation. 
 After inflation\footnote{In principle,
the scalar $\phi$ could itself be the inflaton field as in Ref. \cite{Cline:2019fxx}, but this is not
required for our scenario. Exploring this connection further
is beyond the scope of this work.},
$\phi$ starts rolling at $t= t_i$, 
and the $\phi$ baryon asymmetry $n_\phi(t) = B_\phi r^2 \dot\theta$
evolves according to 
\be
\label{nb-eq}
\frac{1}{a^3}\frac{ \partial}{\partial t} (  a^3 n_\phi ) = -  B_\phi \frac{\partial V}{\partial \theta},
\ee
where $a$ is the Friedman Robertson Walker scale factor and 
we assume a symmetric initial condition, $n_\phi(t_i) = 0$.
Integrating \Eq{nb-eq} approximately yields 
\cite{Rubakov:2017xzr}
\be
n_\phi(t) 
\approx -  \frac{B_\phi}{ {\cal H}(t_i) }\frac{a(t_{i})^3}{a(t)^3} \frac{\partial V}{\partial \theta}(t_i),
\ee
where ${\cal H}  = \dot a/a$ is the Hubble rate and this expression gives the baryon number stored in $\phi$ until it decays through baryon conserving interactions to transfer the asymmetry to the two sectors. Since the baryon density in each sector is set by the corresponding $\phi$ branching fraction, the baryon-to-entropy ratios $\eta^{(\prime)}_{b} \equiv n^{(\prime)}_{b}/
s^{(\prime)}$ satisfy
\be
\label{etas}
\frac{\eta_{b^\prime}}{\eta_{b}} \approx 
\frac{\rm BR_{\phi \to SM^\prime}}{\rm BR_{\phi \to SM}}  \frac{g_{\star, s}(T_{\rm RH})}{g_{\star, s}^\prime(T_{\rm RH}^\prime) }\brac{T_{\rm RH}}{T_{\rm RH}^\prime}^3~,
\ee
where $g_{\star, s}^{(\prime)}$ is the number entropic degrees of freedom in each sector 
and $T^{(\prime)}_{\rm RH}$ is the visible (hidden) reheating temperature.
Thus, the baryon asymmetry transferred to each sector will differ based on model parameters and 
initial conditions in this framework.

To calculate the branching ratios, we postulate baryon conserving interactions of the form
\be
\label{lint}
{\cal L}_{\rm int} =  \frac{\phi}{\Lambda^n} \left(   H^2   \hat {\cal O} + 
H^{\prime^2}\hat {\cal O}^\prime \right) + \mathrm{h.c.}~,
\ee
where $H^{(\prime)}$ is the visible (hidden) Higgs doublet, $\hat {\cal O}^{(\prime)}$ is an operator with compensating baryon number $-B_\phi$,
$\Lambda$ is the cutoff scale of the effective interaction, and $n$ is an integer chosen to ensure that the full expression 
has mass dimension $4$.
The form of the 
operator in \Eq{lint} is schematic, and a non-trivial contraction of $SU(2)_L$ indices may be required to 
ensure the leading $\phi$ coupling is proportional to $v^2$ or $v^{\prime\,2}$; for our purposes, any operator will suffice as 
long as the coefficient preserves this proportionality.
The $\phi$ branching ratios to each sector satisfy 
\be
\label{branch}
\frac{\rm BR_{\phi \to SM^\prime} }{\rm BR_{\phi \to SM}} \approx  \left( \frac{v^\prime}{v} \right)^4~~.
\ee
This approximate expression neglects the contributions from loop level decays through virtual
Higgs propagators that need not be proportional to the same powers of $v$; however, such process are suppressed 
by loop factors of order $(16 \pi^2)^{-2} \approx 4 \times 10^{-5}$ and can be safely neglected. 

Assuming instantaneous reheating through $\phi$ decays, the energy density of each sector is 
proportional to the corresponding branching fraction, so the reheating temperatures satisfy
\be
\label{temp}
\frac{T_{\rm RH}^\prime}{T_{\rm RH}} \approx  \frac{v^\prime}{v} \left[ \frac{g_{\star}(T_{\rm RH})}{g_{\star}^\prime(T_{\rm RH}^\prime)} \right]^{1/4}   , 
\ee
where $g^{(\prime)}_\star$ is the number of relativistic degrees of freedom in each sector. Thus, using 
Eqs.~ (\ref{branch}) and (\ref{temp}), \Eq{etas} becomes
\be 
\label{yields}
\frac{\eta_{b^\prime}}{\eta_b} \approx  \frac{v^\prime}{v}
 \left[ \frac{g_{\star}^\prime(T_{\rm RH}^\prime)}{g_{\star}(T_{\rm RH})} \right]^{1/4}~,
\ee
yielding a simple relationship between the asymmetries of our sectors. Note that in the
$g_\star = g_\star^\prime$ limit, \Eqs{temp}{yields} imply that all energy density ratios are equal in the two sectors (e.g., $\rho_b/\rho_\gamma 
= \rho_{b^\prime}/\rho_{\gamma^\prime}$), as required to approximate the symmetry in \Eq{sym}.

Since the hidden sector satisfies $m_{e^\prime} \ll m_{p^\prime}$, the binding energy of Hydrogen in
both sectors obeys 
\be
\label{BH}
B^{(\prime)} = \frac{\alpha^2}{2} \mu^{(\prime)} \propto v^{(\prime)},
\ee
where $\mu^{(\prime)}$ is the electron-proton reduced mass. Therefore, for $g_\star = g_\star^\prime$ in \Eq{temp}, we predict $B^\prime/T^\prime \approx  B/T$, which suffices to trigger hidden recombination around $z \approx 1100$.
In order to ensure that $T^\prime_{\rm RH} /T_{\rm RH} < 1$, we require $v^\prime  < v$, so massive elementary particles are uniformly lighter in the hidden sector. Note that the expression in \Eq{BH} is only an approximate equality because the redshift of recombination is logarithmically sensitive to $\eta_b$ and our scenario predicts $\eta_b \ne \eta_{b^\prime}$ from \Eq{etas}. However, since the temperature ratio
of the two sectors is only of order one \cite{Cyr-Racine:2021alc}, this mild deviation is negligible for our purposes. 

Finally, we note that our model is more predictive than the phenomenological study in Ref. \cite{Cyr-Racine:2021alc} because the 
ADM fraction of the total dark matter density is 
\be
\label{fadm}
f_{\rm adm} \equiv \frac{\rho_{\rm adm}}{\rho_{\rm cdm}} = \frac{\rho_{b}}{\rho_{\rm cdm} }
\brac{v^\prime}{v}^4 
\approx 0.05 \brac{T^\prime/T}{0.7}^4 \!\!, ~~
\ee
where we have assumed $\rho_{b}/\rho_{\rm cdm} \approx 1/5$ and used \Eq{temp} with $g_\star = g^\prime_\star$.\footnote{Note that the precise value of $f_{\rm adm}$ needs to be determined self-consistently by varying $v'/v$, $\rho_b$ and $\rho_\mathrm{cdm}$ simultaneously while fitting to cosmological data.\label{ft:fadm_val}} 
Thus $f_{\rm adm}$ is correlated with the temperature ratio and lies naturally in the range favored to reconcile early- and late-time measurements of $H_0$ for the best fit value $T^\prime/T \approx 0.7$ in Ref. \cite{Cyr-Racine:2021alc}.

\section{Caveats and Comments}

\subsection*{Affleck-Dine Mass Scale}

Assuming $\phi$ decays take place during a cold, matter-dominated phase, both sectors are 
in the broken electroweak phase throughout reheating; if this were not the case, the branching fractions would not 
necessarily scale according to the relation in \Eq{branch}. Furthermore, in the broken vacuum, the $H^2$ proportionality in 
\Eq{lint} can be expanded using $H = [0, (v+h)/\sqrt{2}]^T$ to generate interactions of the form $v h \phi \,\hat {\cal O}/\Lambda^n$, whose branching fraction scales as $\propto v^2$, not $v^4$, as desired in \Eq{branch}. Such decays can be kinematically forbidden if $m_\phi < m_{h^\prime}$.

\subsection*{Reheat Temperature}
To ensure that the branching ratios in
\Eq{branch} are satisfied throughout the early universe, we demand that
both sectors reheat to temperatures below the scale of electroweak symmetry breaking, $T_{\rm RH}^{(\prime)} \lesssim v^{(\prime)}$. Since this requirement necessarily implies that some fields 
in each sector will not be produced, we must nonetheless ensure that $g_\star(T_{\rm RH}) = g^\prime_\star(T^\prime_{\rm RH})$.
However, since the field content in the two sectors is identical, this can be achieved across a wide range of temperature 
ratios. For example, with $v^\prime/v = 1/2$ and $T_{\rm RH} =$ 50 GeV, all leptons, light quarks, and massless
gauge bosons are produced in both sectors, but neither thermalizes its $W^\pm,Z^0, h$, or $t$ particles.

Assuming a non-relativistic $\phi$-dominated universe, instantaneous $\phi$ decays, and rapid equilibration (compared to Hubble expansion) in each sector, the reheating temperature of each sector can be approximated as
\be
\label{rh}
~~ T_{\rm RH}^{(\prime)} 
   \approx
  1\, {\rm GeV} \brac{ \rm  BR_{\phi \to SM^{(\prime)} }     }{0.33}^{1/4} \! \brac{0.5\, \mu \rm s}{ \tau_\phi}^{1/2} 
  \hspace{-0.2cm} \!\!\! ,~~~~
\ee
where $\tau_{\phi}$ is the scalar lifetime. However, the scaling in \Eq{rh} is highly model dependent and can be modified, for example, 
with additional decay channels for $\phi$ or by parametric resonance effects \cite{Kofman_1997}. Our scenario is compatible
with any of these reheating variations as long as $\phi$ decays in the broken electroweak vacuum and the relation in
\Eq{branch} is preserved to good approximation.

\subsection*{Choosing The Decay Operator}

To realize the branching ratio relation in \Eq{branch}, the operators in \Eq{lint} must be chosen 
with care. Since the early universe is always in the broken electroweak phase, $\phi$ decays must 
directly generate a net baryon asymmetry; a purely lepton number asymmetry would not
yield a baryon asymmetry here since sphalerons are always out of equilibrium in our scenario. 

Furthermore, since SM operators with net baryon number involve many insertions of quark fields (e.g., $ {\hat {\cal O}} = u^c d^c d^c$ where 
$u^c$ and $d^c$ are respectively up- and down-type $SU(2)_L$ singlet quarks), the exponent $n$ in \Eq{lint} is a large integer. This suppression makes it generically difficult
to reheat the universe above the MeV scale while keeping $m_\phi < m_h$, required to maintain the relation in \Eq{branch}. However, in the presence
of additional baryon-charged fields in each sector this problem can be avoided. 

As a toy example, we can add to each sector a gauge singlet 
Weyl fermion $\chi^{(\prime)}$ and its Dirac partner $\chi^{c  (\prime)}$ with
 baryon number $ \mp B_\phi/2$, respectively. This enables us to posit
  $\phi$ decay interactions of the form
 $\phi |H| ^2 \chi \chi/\Lambda^2$, which can ensure the relations in \Eq{branch} while giving $\phi$ sufficiently
 prompt $\phi \to \chi^{(\prime)} \chi^{(\prime)}$ decay channels. Once the universe is populated with $\chi$ particles at $T_{\rm RH} >$ MeV,
 prompt $\chi \to 3q$ decays can proceed through a $\chi^c u^c d^c d^c/M^2$ operator
 \footnote{It is generically easy to ensure $\tau_\phi \gg \tau_\chi$, where $\tau_\chi \sim M^4/m_\chi^5 \sim 10^{-9}\; {\rm s}\; (M/100 \, {\rm TeV})^4(10 \, {\rm GeV}/m_\chi)^5$ is
 the $\chi$ lifetime. Thus, $M$ can be sufficiently large to evade empirical bounds.
 }
 to transfer the baryon asymmetry to SM particles, where $M$ is the mass of a heavy particle that has been integrated out. While this
 realization satisfies all of our requirements, the gauge and baryon charge assignments for $\chi$ also allow direct $\phi \chi \chi$ couplings, which 
would induce decays that violate the VEV scaling in \Eq{branch}. This issue can be avoided if  $\phi$ carries baryon minus lepton number, which allows
the operator $\phi (LH)^2 \chi\chi/\Lambda^5$ and forbids the renormalizable $\phi \chi \chi$ interaction\footnote{Forbidding the $\phi \chi\chi$ operator but 
allowing the $\phi (LH)^2 \chi \chi/\Lambda^5$  interaction may require baryon minus lepton number to be gauged and spontaneously broken at low energies. 
}; as noted 
below, this operator dimension can still yield a sufficiently high reheating temperature for BBN. 

For all candidate decay operators, it is important to forbid mixed terms that involve fields from both sectors.
From the example above, a complete ultraviolet model must forbid operator variations of the form $\phi  (LH) \chi (L^\prime H^\prime) \chi^\prime /\Lambda^5$,
which would spoil the required relation in \Eq{branch}. Such a sequestration can be achieved, for example, in extra dimensional models 
in which the two sectors live on different four dimensional branes, but $\phi$ lives in a higher dimensional bulk, so hybrid 
operators involving both sectors are forbidden by locality (see Refs. \cite{Csaki:2004ay,Sundrum:2005jf} for examples). 
The mixed operators 
can also be suppressed if $\chi^{(\prime)}$ is endowed with a parity symmetry under which $\chi \chi^\prime$ transforms non-trivially; this symmetry is then spontaneously broken to allow for $\chi^{(\prime)}$ decays to baryons.

\subsection*{Avoiding Thermalization}
\begin{figure*}[h!]
  \centering
  \includegraphics[width=0.9\textwidth]{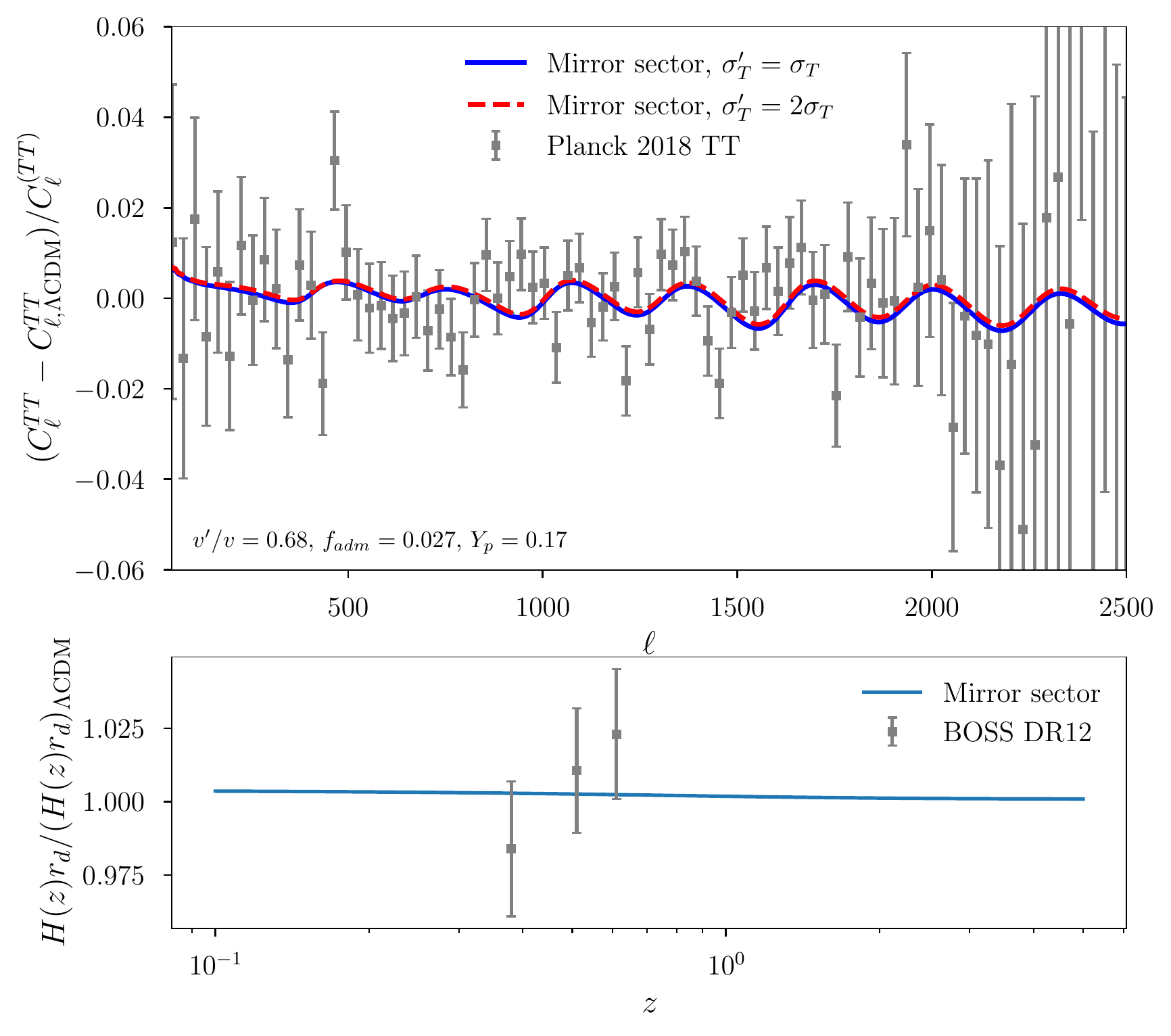}
  \caption{Impact of a mirror sector on representative CMB (upper panel) and BAO (lower panel) observables. 
  The upper panel shows the relative shift of the TT power spectrum for two mirror sector models: one with the dark Thomson cross-section $\sigma_T^\prime$
  equal to the SM one, $\sigma_T$ (as assumed in Ref.~\cite{Cyr-Racine:2021alc}), and one where $\sigma_T^\prime \approx 2\sigma_T$ as predicted in our set-up. The differences are negligible and both models are compatible with the observed spectra, whose binned residuals relative to $\Lambda$CDM are shown by the points with error bars~\cite{Planck:2019nip}. The mirror sector models have $H_0 \approx 73.4$ km/s/Mpc, $f_\mathrm{adm} = 0.027$, $T'/T = 0.68$, and $Y_p = 0.17$ following Ref.~\cite{Cyr-Racine:2021alc} (the other cosmological parameters are given in the text).
  In the lower panel we show $H(z) r_d$ relative to its $\Lambda$CDM value, where $r_d$ is the sound horizon at the drag epoch, along with BAO measurements of this quantity from BOSS~\cite{BOSS:2016wmc}. 
Note that $\Lambda$CDM and the mirror sector models are nearly indistinguishable in this observable because of the scaling symmetry $H\to fH$, $r\to r/f$ that the mirror sector implements~\cite{Cyr-Racine:2021alc}.
\label{fig:cmb_and_bao}}
\end{figure*}

Since the two sectors in our setup must not thermalize with each other, we conservatively 
demand that $\phi$ never thermalize with the SM whose energy density
is always greater. For an interaction rate based on \Eq{lint} in the broken electroweak phase, 
$\Gamma_{\rm \phi-SM} \sim  v^4 T_{\rm RH}^{2n-3}/\Lambda^{2n} < \cal H$ at reheating, so the suppression scale must satisfy 
 \be
 \Lambda \gtrsim \left(  \frac{m_{\rm Pl}  v^4 T_{\rm RH}^{2n-5}}{ \sqrt{ g_\star(T_{\rm RH}) } }\right)^{1/2n}
 \!\!
 \approx   2\,{\rm TeV}\brac{T_{\rm RH}}{10 \, \rm GeV}^{1/2}\!\!\!,~~~~
 \ee
where $m_{\rm Pl} = 1.22 \times 10^{19}$ GeV is the Planck mass and in the last step we took $n=5$.
Thus, one can ensure that $T_{\rm RH} \gtrsim$ MeV as required 
by the success of standard BBN~\cite{Kawasaki:2000en,Hannestad:2004px,deSalas:2015glj,Hasegawa:2019jsa} without
thermalizing $\phi$ with either sector. 
We expect that a careful treatment of the phase-space in the 
above thermalization rate would only change the bound on $\Lambda$ by $\mathcal{O}(1)$ because of the $2n$'th root.

In addition to avoiding thermalizing the two sectors through the $\phi$ interactions, we must also 
ensure that mixed operators of the form $\lambda |H|^2 |H^\prime |^2$ are either forbidden
or sufficiently suppressed such that $ hh \leftrightarrow h^\prime h^\prime$ reactions are 
always slower than Hubble expansion. As noted above, such hybrid interactions
can be naturally suppressed by locality in extra dimensional theories
that require the two sectors to live on different branes.

\section{Discussion}

We have introduced a realistic model of atomic
dark matter that brings local and CMB measurements of $H_0$ into concordance following 
the phenomenological study in Ref. \cite{Cyr-Racine:2021alc}. 
Our approach
is inspired by Twin Higgs models in which a hidden sector contains an identical
copy of the SM, but with a slightly different Higgs VEV, $v'$, and temperature, $T'$.
The baryon asymmetry and reheating temperature in each sector is set by the 
VEV-dependent branching ratio of an Affleck-Dine scalar field $\phi$, and all differences
between sectors are governed by $v^\prime/v$.
Since the $\phi$ branching ratio to the visible and hidden sectors scales as $\propto v^4$ and $v'^4$, respectively,
the hydrogen binding energy-to-temperature ratio, $B/T$, 
is the same in both sectors and recombination occurs for all atomic species
at $z \approx 1100$. As a result, there is no fine tuning required to ensure this coincidence.

Our model features several notable differences with respect to the phenomenological
treatment of Ref. \cite{Cyr-Racine:2021alc}: 

\begin{enumerate}

  \item Because our hidden sector has a smaller electron mass, the Thomson cross section satisfies $\sigma_T^\prime = (v/v^\prime)^2 \sigma_T =2.1 \sigma_T$. So, even if recombination still occurs at $z \approx 1100$ on account of $B/T = B^\prime/T^\prime$, this larger $\sigma_T^\prime$ can change the acoustic oscillations of the hidden sector baryons and thus modify the CMB/BAO observables indirectly. To test the importance of this effect, we implemented a mirror sector model in the Boltzmann code \texttt{CLASS v 3.1.1}~\cite{Lesgourgues:2011rh} by 
      modifying the existing interacting DM module~\cite{Archidiacono:2019wdp}: we simply rescaled the DM-dark radiation interaction terms by the free-electron fraction 
      of the SM (this is a reasonable approximation since fast variations in this quantity, i.e. recombination, occur simultaneously 
  in the visible and dark sectors by construction) and scaled the interaction coefficient, \texttt{a\_idm\_dr}, to match $\sigma_T$. The redshift-dependence of this interaction matches that of Thomson scattering for $n=2$ in the notation of Ref.~\cite{Archidiacono:2019wdp}.\footnote{The interacting DM module contains additional parameters: $\alpha_\ell$ encodes the damping of higher multipole moments of the dark photon distribution -- we take $\alpha_2 = 9/10$ and $\alpha_{\ell>2} = 1$ by matching the notation of Ref.~\cite{Archidiacono:2019wdp} to $\Lambda$CDM photons~\cite{Ma:1995ey}; \texttt{stat\_f\_idr} is set to $1$ for bosonic dark radiation; the interacting DM mass, \texttt{m\_idm} is taken to be $m_p$.} In Fig.~\ref{fig:cmb_and_bao} we show that a different scattering 
      cross-section has very little impact on the quality of the fit to CMB and BAO observables. There we take cosmological parameters motivated by 
      the detailed MCMC analysis in Ref.~\cite{Cyr-Racine:2021alc}: $v'/v = 0.68$, $Y_p =0.17$, $f_\mathrm{adm} = 0.027$; note that other parameters were not explicitly provided in that paper so we picked values that give a reasonable fit by eye: $\Omega_\mathrm{cdm}h^2 = 0.142$, $\Omega_{b}h^2 = 0.0224$, $100\theta_s=1.0425$, 
      $\ln 10^{10} A_s = 3.03$, $n_s = 0.958$, $\tau_{reio} = 0.0539$, and $N_{\mathrm{fs}} = 3.30$, where $N_{\mathrm{fs}}$ is the number of free-streaming degrees of freedom (i.e. SM and mirror neutrinos). The total number of relativistic degrees of freedom, $N_{\mathrm{eff}} \approx 4.24$, includes the contribution of the non-free-streaming dark photons. 
      A similar Boltzmann code implementation of a twin mirror sector was recently studied in Ref.~\cite{Bansal:2021dfh}, but their focus was on models with $v'/v > 1$.
      It would be interesting to perform a detailed cosmological analysis for our specific realization of the mirror sector in a future work -- we expect this would give rise to only small shifts in the preferred values of the cosmological parameters.

\item From \Eq{fadm} our scenario features a one-to-one correspondence between
$f_{\rm adm}$ and the hidden/visible temperature ratio, so our model has fewer
free parameters than \cite{Cyr-Racine:2021alc}. It is also notable that 
the relation in \Eq{fadm} is automatically consistent with the approximate best fit values
$f_{\rm adm} \approx 0.05$ and $T^\prime/T \approx 0.7$ (see also footnote~\ref{ft:fadm_val}).

\item Unlike Ref. \cite{Cyr-Racine:2021alc}, which assumed $\eta_{b^\prime} = \eta_b \approx 9 \times 10^{-11}$ \cite{ParticleDataGroup:2020ssz}, 
from \Eqs{etas}{yields}, our framework predicts a different baryon asymmetry in the hidden sector. Since we, nonetheless, recover
 $\rho_{b^\prime}/\rho_{\gamma^\prime} = \rho_b/\rho_\gamma$, this difference does not affect any CMB observables, but can play an
 important role in hidden sector BBN. While hidden BBN does not affect visible sector CMB observables, it may have interesting observational consequences
 that warrant further study \cite{Krnjaic:2014xza,Hardy:2014mqa,Detmold:2014qqa,Redi_2019,Mathur:2021gej}.

\end{enumerate}

As noted in \cite{Cyr-Racine:2021alc}, this framework is in
generic tension with both the direct measurement of the primordial helium
fraction and the theoretical prediction of Big Bang nucleosynthesis accounting for the large value of $\Delta N_{\rm eff}\approx 1.6$ (assuming
the full SM-like field content in the hidden sector). 
BBN with such a large enhancement to the 
expansion rate predicts $Y_p\approx 0.266\pm 0.0053$, higher than in the standard cosmology with $Y_p\approx 0.247\pm 0.0046$ (with $n_b/n_\gamma = 6.13\times 10^{-10}$~\cite{Planck2020});\footnote{These predictions depend on the specific BBN code and reaction network used, but the discrepancy is unaltered. We used \texttt{AlterBBN} v 2.1~\cite{Arbey:2018zfh} to estimate the theoretical uncertainties using a Monte Carlo procedure.} thus the theoretical prediction of $Y_p$ within this model appears to be even more inconsistent with the CMB inference of the same quantity. Even if the hidden sector is populated after BBN, through, e.g., the decay of a non-relativistic particle, the CMB-preferred value still disagrees both with the direct late-universe measurement and the (now) standard $Y_p$ yield from BBN.
These points of tension persist in our scenario, but because we approximately realize the scaling
relation in \Eq{sym}, we also inherit a good fit to CMB observables and a larger value of 
$H_0 \approx 73$ km s$^{-1}$Mpc$^{-1}$, provided that $Y_p$ and $\Delta N_{\rm eff}$ are allowed to float.

It would be interesting to study whether hidden sector model variations can overcome these observational
limitations. For example, it may be possible to realize our scenario with only a subset of SM generations
in the hidden sector, in analogy with Fraternal Twin Higgs models \cite{craig2015naturalness}, but such
studies are beyond the scope of this work. 

Finally, we note that our scenario may imply several interesting consequences that are worth exploring in 
detail. For example, an atomic dark sector with identical field content may yield dark nuclei \cite{Krnjaic:2014xza,Hardy:2014mqa,Detmold:2014qqa}, galactic disks \cite{Fan:2013yva,McCullough:2013jma}, and stars \cite{Hippert:2021fch,Sandin:2008db,Curtin:2019lhm,Curtin:2019ngc,Curtin:2020tkm,Winch:2020cju}
but exploring the observational implications of these structures is beyond the scope of this work. 
Moreover, UV completions of the various effective operators required to realize our proposal might lead to additional signatures.
For example, it may be possible for neutral hidden-visible oscillations between each sector's neutrinos, neutrons, and photons, as long as the mixing interactions that enable these processes do not thermalize the two sectors at early times.

\bigskip
\begin{acknowledgments}  
\textbf{Acknowledgments.} 
We thank Asher Berlin, Kimberly Boddy, David Curtin, Raymond Co, Jeff Dror, 
Xiao Fang, Paddy Fox, Roni Harnik, Dan Hooper, Julian Mu\~noz, Harikrishnan Ramani, Dan ``Danimal" Scolnic, and Flip Tanedo for helpful conversations.
This work is supported by the Fermi Research Alliance, LLC under Contract No. DE-AC02-07CH11359 with the U.S. Department of Energy, Office of Science, Office of High Energy Physics. NB was supported in part by NSERC, Canada.
This work was partly completed at the Aspen Center for Physics, which is supported by National Science Foundation grant PHY-1607611.
\end{acknowledgments}

\bibliography{ADM_H0}

\end{document}